\documentclass[%
 reprint,
 amsmath,amssymb,
 aps,
]{revtex4-2}

\usepackage{babel,blindtext}
\usepackage{tipa}
\usepackage[utf8]{inputenc}
\usepackage{siunitx}
\usepackage{bm}
\usepackage{xcolor}
\usepackage{graphicx}% Include figure files
\usepackage{dcolumn}% Align table columns on decimal point
\usepackage{bm}% bold math
\begin{document}

\preprint{APS/123-QED}

\title{Comprehensive deep learning model for optical holography}

\author{Alim Yolalmaz$^{1, 2, 3}$ and Emre Y\"{u}ce$^{1, 2, 3,}$}
\email{eyuce@metu.edu.tr}
 
\affiliation{
 $^{1}$Programmable Photonics Group, Department of Physics, Middle East Technical University, 06800 Ankara, Turkey\\
 $^{2}$Micro and Nanotechnology Program, Middle East Technical University, 06800 Ankara, Turkey\\
 $^{3}$The Research and Application Center for Space and Accelerator Technologies (ODTÜ-\"{I}VMER), Middle East Technical University, 06800 Ankara, Turkey}%

\date{\today}% It is always \today, today,

\begin{abstract}

Holography is a vital tool used in various applications from microscopy, solar energy, imaging, display to information encryption. Generation of a holographic image and reconstruction of object/hologram information from a holographic image using the current algorithms are time-consuming processes. Versatile, fast in the meantime accurate methodologies are required to compute holograms performing color imaging at multiple observation planes and reconstruct object/sample information from a holographic image for widely accommodating optical holograms. Here, we focus on design of optical holograms for generation of holographic images at multiple observation planes and colors via a deep learning model the DOENet. The DOENet produces optical holograms which show multitasking performance as multiplexing color holographic image planes by tuning holographic structures. Furthermore, our deep learning model retrieves an object/hologram information from an intensity holographic image without requiring phase and amplitude information from the intensity image. We show that reconstructed objects/holograms show excellent agreement with the ground-truths. The DOENet does not need iteratively reconstruction of object/hologram information while conventional object/hologram recovery methods rely on multiple holographic images at various observation planes along with the iterative algorithms. We believe that the DOENet based object/hologram reconstruction and generation of holographic images may speed up wide-area implementation of optical holography.

\vspace{0.3cm}

\end{abstract}

\maketitle

%\tableofcontents

\section{\label{sec:level1}Introduction}

Optical holography is a superior tool to retrieve phase and amplitude of light from an intensity image, which bears detailed information of the object as size, shape, and refractive index. Reconstructed object/hologram from an intensity image has important roles in high-security encryption\cite{Fang2020, Javidi2000a}, microscopy\cite{Liebel2020}, data storage\cite{Yoneda2019}, 3D object recognition\cite{Javidi2000}, and planar solar concentrator\cite{Yolalmaz2020, Guen2021, Yolalmaz2021}. Besides the information gained from intensity images the process is reversible; the phase and amplitude information can be used to generate intensity images for numerous applications such as imaging\cite{Zheng2015, Rivenson2017,Leonardo2010}, photostimulation\cite{Anselmi2011}, printing\cite{Kim2015}, optical beam steering\cite{Hayasaki1999}, aberration correction\cite{Love1997}, display\cite{Moon2014,Zhao2018}, and augmented reality\cite{Hong2013}. Optical holography enables image formation at different observation planes or through a sample without requiring any focusing elements or mechanical scanning. For generation of a holographic image fine-tuned phase/amplitude distribution of a hologram is required that increases design duration of a hologram. Considering wide-area implementation of optical holography for both retrieving object/hologram information and generation of a holographic image a versatile methodology with a short design/optimization duration is highly demanded.

For generation of optical holographic images and object/hologram recovery, there are a variety of algorithms frequently used\cite{Zhang2017, Kettunen2004, Gerchbergpace1972, Liu1987, Fienup1982, Kim2012}. These algorithms are easy to employ and yield well performance but require iterative optimization. Furthermore, strong push to convergence for tolerable error may cause the algorithms to yield physically infeasible patterns. In contrast to these algorithms, deep learning correlates an intensity distribution to a hologram without reconstruction of phase and amplitude information from an intensity distribution thanks to its data-driven approach. Deep learning is a superior tool that presents important achievement especially in holography for imaging\cite{ Wu2018, Jaferzadeh2019a, Ning2020, Schnell2014}, microscopy\cite{Pitkaeaho2019, Liu2020, Liu2019, Luo2019a}, optical trapping\cite{David2018}, and molecular diagnostics\cite{Kim2018}. However, for generation of optical holograms which provide holographic images at different observation planes and wavelengths, versatile neural networks are required\cite{Horisaki2018, Eybposh2020, Lee2020a}. This issue is weakly addressed in the literature, and proper modalities are demanded for generation of optical holographic images having diverse properties: different observation planes, wavelengths, and figures. Using a comprehensive neural network model, design of holograms could be accelerated that leads to generation of optical holography at different observation planes and wavelengths.

In this study for the first time to our knowledge a single deep learning architecture is employed to generate (i) a multi-color three images at a single observation plane, (ii) a single color three images at multiple observation planes, (iii) a multi-color three images at multiple observation planes. In addition to generation of holographic images, the deep learning model provides retrieval of object/hologram information from an intensity image without requiring phase and amplitude information of an intensity holographic image thanks to statistically retrieval behavior of deep learning. Moreover, we obtain holographic images with twin-image free by eliminating phase and amplitude retrieval. Training the deep learning model with a data set which is a one-time process lasting less than an hour speeds up generation of holographic images and object/hologram recovery from an intensity image down to two seconds.

\section{Methods}

\subsection{Data set generation}

A computer-generated hologram (CGH) could be a phase mask or a three-dimensional object where thickness value changes spatially. Thickness distribution of a CGH is interpreted also in a phase distribution with refractive index value of a holographic plate and wavelength of light. With the angular spectrum method (ASM), intensity distribution of a holographic image at each wavelength of a light source is computed by considering thickness distribution of a CGH. Using Eqs. (1-3) standing for Fresnel-Kirchhoff diffraction integral, an electric field of light propagates from a hologram plane to a holographic image plane is calculated. The electric field of light at the hologram plane $U_{Hologram}$ is obtained with an incident amplitude of light $A_{incident}$, thickness distribution of the hologram $t_{Hologram}$, refractive index of the hologram material $n$, and wavelength of light source $\lambda$ after utilizing Eq. 1. p and q in Eq. 1 are indices of spatially varying amplitude of light at the hologram plane. The light wave from the hologram plane transforms into the image plane with a kernel transformation function $G$ in Eq. 2. The kernel transformation function has light propagation parameters as an observation plane distance from the hologram plane to the image plane $d$, location of pixels at the hologram plane (x,y), and the image plane (X,Y). a and b are indices at the image plane. Later, we obtain amplitude of light at the image plane $U_{Image}$ with Eq. 3. The intensity of the holographic image $I_{Image}$ is attained with a square of the light amplitude at the image plane $U_{Image}$.

We use correlation coefficient $\rho$ as a metric to evaluate similarity between a ground-truth/ideal holographic image $I_{Ideal}$ and a designed holographic image $I_{Designed}$ with the deep learning model/ASM (see Eq. 4). The same metric evaluates similarity between a ground-truth hologram $I_{Ideal}$ and a reconstructed hologram with the deep learning model $I_{Designed}$. N in Eq. 4 is number of pixels in the ideal image and the designed image; $I_{Ideal, i}$ and $I_{Designed, i}$ in Eq. 4 are intensity values of ith pixel of the ideal image and the designed image, respectively. The other terms in Eq. 4 are mean of the ideal image $\mu _{{I_{Ideal} }}$, mean of the designed image $\mu _{{I_{Designed} }}$, standard deviation of the ideal image $\sigma_{I_{Ideal}}$, and standard deviation of the designed image $\sigma_{I_{Designed}}$. We compute a correlation coefficient for a ground-truth/ideal holographic image $I_{Ideal}$ and corresponding designed holographic image $I_{Designed}$ at each observation plane distance $d$ and each wavelength of light $\lambda$.

\begin{widetext}
\begin{equation}
{U_{Hologram}(p,q,\lambda)=A_{incident}(p,q,\lambda)*\text{exp}({2\pi j *t_{Hologram}(p,q)*\lbrack n(\lambda)-1 \rbrack/\lambda})},
\end{equation}

\begin{equation}
{G(p,q,a,b,\lambda,d)=\sum_{pq}\left ( \frac{1}{j\lambda d} \right )*\text{exp}\left ( \frac{j2\pi d}{\lambda} \right )*\text{exp}\left [j\pi\frac{\left \{  (y_{pq}-Y_{ab})^2+(x_{pq}-X_{ab})^2  \right \}}{\lambda d}  \right ],}
\end{equation}

\begin{equation}
{U_{Image}(a,b,\lambda,d)=\sum_{pq}U_{Hologram}(p,q,\lambda)*G(p,q,a,b,\lambda,d)},
\end{equation}

\begin{equation}
{\rho(I_{Ideal},I_{Designed})=\frac{100}{N-1}\sum_{i=1}^{N}\frac{\left ( I_{Ideal, i} - \mu _{{I_{Ideal} }} \right  ) * \left ( I_{Designed, i} - \mu _{{I_{Designed} }} \right  )}{\sigma_{I_{Ideal}} * \sigma_{I_{Designed}}} }.
\end{equation}

\end{widetext}

\begin{figure*}[!htb]
    \centering
        \includegraphics[width=150 mm]{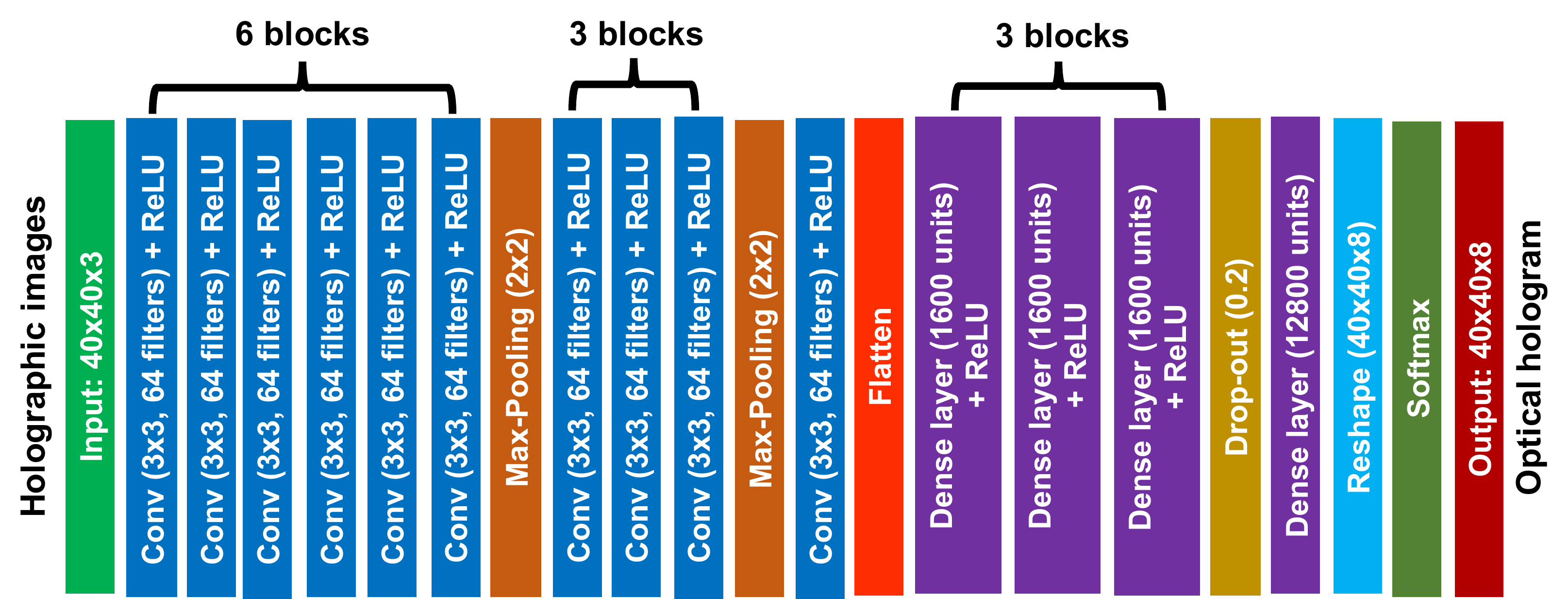}
    \caption{The DOENet architecture. Each color represents a different data operation. The model takes holographic images with a size of 40-by-40-by-3, performs feature extractions through several CNN layers, and correlates holographic images to holograms with a size of 40-by-40-by-8 by using detected features.}
    \label{model}
\end{figure*}

Using spatially varying 8-level thickness profile of a hologram, we tune phase of a light source to form a holographic image. The holographic image is designed with a linearly polarized continuous light at normal incidence. Resolution of a hologram and a holographic image are 40-by-40. Each pixel of a hologram has a thickness value ranging from 1 $\mu m$ to 8 $\mu m$, which is divided into 8 equal discrete steps. We evaluate performance of a hologram with a correlation coefficient $\rho$ in terms of percent with Eq. 4 during organizing thickness distribution of the hologram with the local search optimization algorithm. The local search optimization algorithm sequentially changes thickness of each pixel in the hologram after generating a random hologram distribution to form desired intensity images at selected frequencies of the light and observation plane distances. We used two decision-making criteria: AND and MEAN, which are also called logic operations, in the algorithm to minimize difference between an ideal image (uniform and binary image distribution) $I_{Ideal}$ and a designed image $I_{Designed}$. With AND logic operation, we organize a hologram thickness distribution considering increases in values of the correlation coefficient at each wavelength and each observation plane simultaneously. With MEAN logic operation, the hologram thickness distribution is tuned when average correlation coefficient computed with correlation coefficients at all wavelengths and all observation plane increases. At the end of designing a hologram, we obtained 51200 optical hologram distributions and corresponding holographic images at each light frequency/observation plane.

\subsection{Deep learning model}

In this manuscript, we employed the deep learning model the DOENet presented in our previous work\cite{Yolalmaz2021a} (see Fig \ref{model}). The DOENet demonstrates reconstruction of an object thickness distribution/hologram by using holographic intensity images at all wavelengths of light and observation planes. The framework uses holographic intensity images to train weights of the model and understands how to transform spectral and spatial information of incident light encoded within holographic images to optical holograms. The DOENet receives multiple images at different observation planes/colors as inputs and produces a hologram in terms of thickness that can reconstruct input images at predefined depths/colors. The model takes three-channel intensity distributions two of which belong to spatial size of intensity distributions, and the third is for either different wavelengths, observation planes, or wavelengths plus observation planes. The DOENet consists of 10 convolutional neural network (CNN) layers with 64 filters and a filter size of 3-by-3, two max-pooling layers with a size of 2-by-2, a flattening layer, three dense layers with 1600 units, one dense layer with 12800 units, a drop-out layer with a factor of 0.2, a reshaping operation, and a softmax activation function. For tuning weights of the model, we utilize 51200 holographic images at each frequency channel/observation plane and corresponding 8-bit thickness distribution of holograms as a training data set. Before tuning weights of the DOENet, we convert hologram information to a one-hot vector with eight classes for obtaining better performance with discrete hologram distributions. During training with 90\% of all data, the DOENet minimizes categorical cross-entropy loss function. The DOENet iteratively updates the model’s weights and biases using the adaptive moment estimation (Adam) optimizer with a learning rate of $10^{-4}$ during back-propagation. The model is implemented using TensorFlow, an open-source deep-learning software package. The training and blind testing of the network were performed on a workstation with 32 GB RAM and an NVIDIA Quadro P5000 GPU. The training process lasts less than an hour for 50 epochs. Once the training was completed at each epoch, we test our model with a validation data set which is 10\% of the input data set. After the training, the network inference time for reconstruction of an object/hologram distribution with reversing one-hot vector operation is almost two seconds on average.

\section{Results and discussion}

\begin{figure}[!htb]
    \centering
    \includegraphics[width=80 mm]{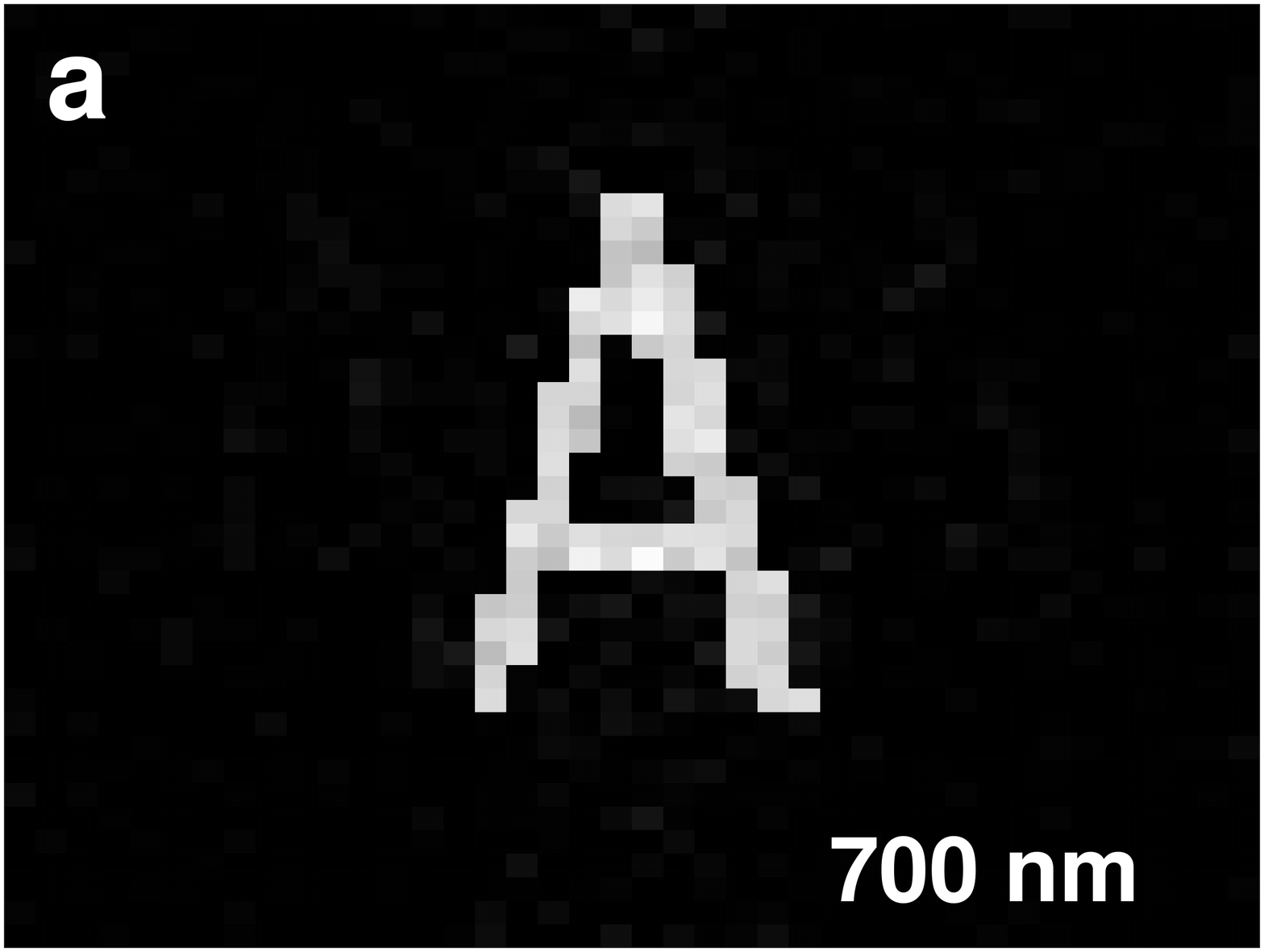}\\
        \includegraphics[width=80 mm]{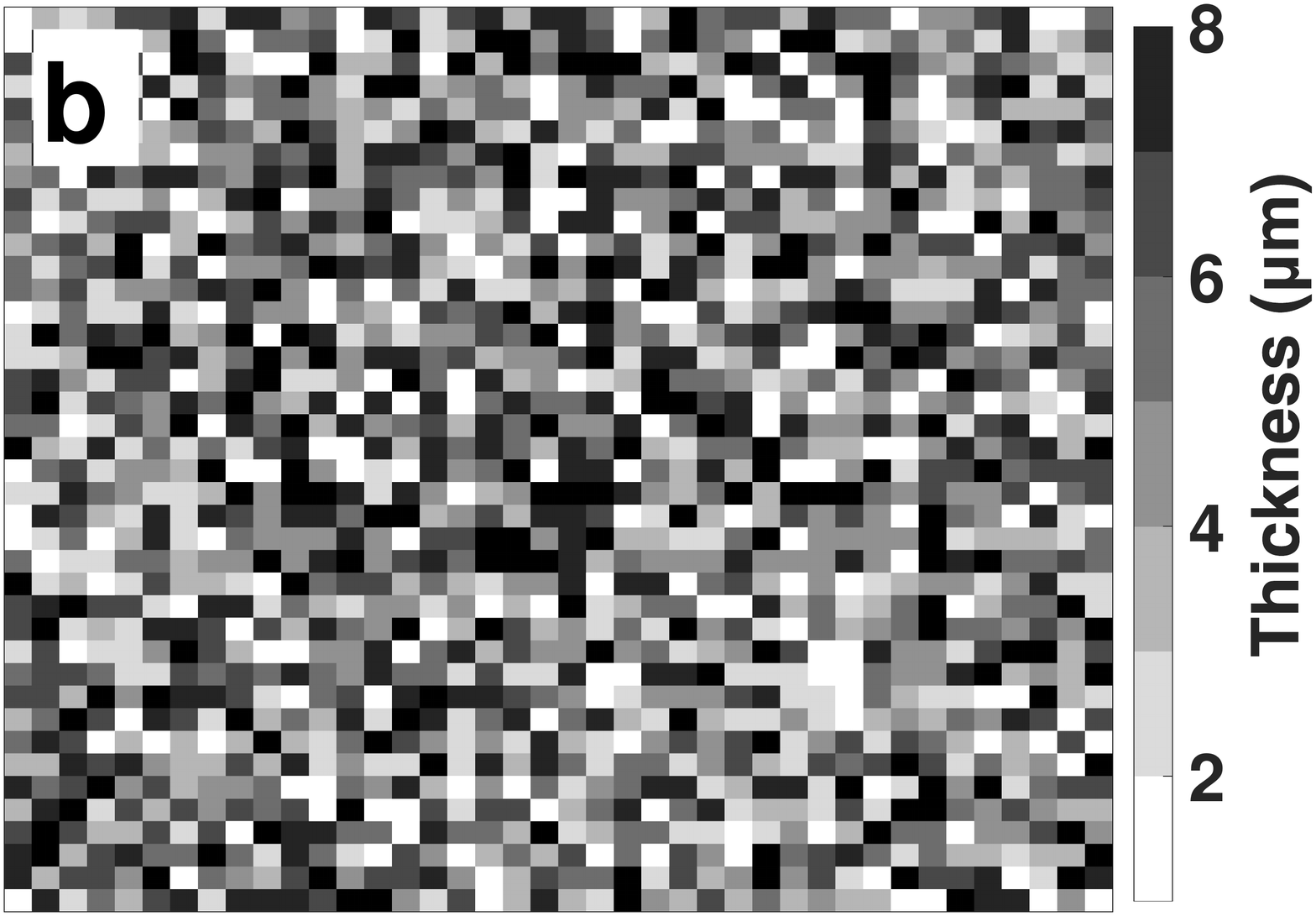}\\
            \includegraphics[width=80 mm]{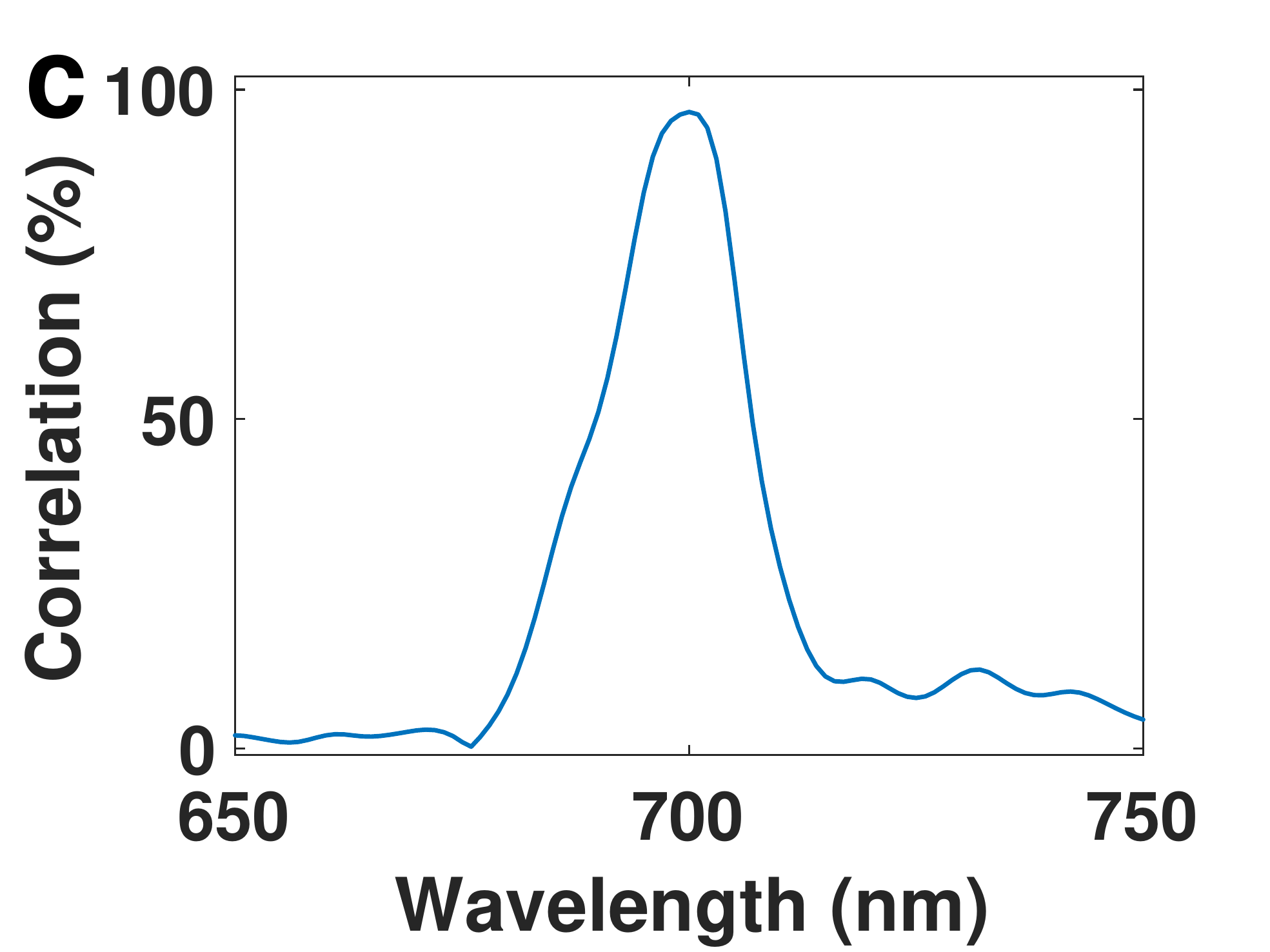}
    \caption{(a) The intensity image of letter A; (b) Thickness distribution of designed CGH which produces the image as seen in (a); (c) The correlation spectrum of holographic images at wavelengths between 650 nm - 750 nm obtained with the hologram presented in (b).}
    \label{wavelengthselection}
\end{figure}

The selection of design wavelengths is a crucial step to form holographic intensity images at the image plane without a cross-talk between images at different frequencies of light. For this concern, at first, we inspect spectral bandwidth which a holographic image presents when a hologram designed for this image is illuminated by broadband light. We design a CGH that images an intensity distribution of letter A at a wavelength of 700 nm as seen in Fig. \ref{wavelengthselection}a. This image provides a correlation coefficient of 96.6\% with a uniform and binary image of letter A after utilizing Eq. 4. As seen in this image, we generated a clear and almost uniform intensity distribution of letter A which is obtained with a hologram distribution having a thickness profile in Fig. \ref{wavelengthselection}b. When the same CGH is illuminated by a uniform broadband light (650 nm - 750 nm), the correlation coefficient varies through different illumination wavelengths of light as seen in Fig. \ref{wavelengthselection}c, and correlation coefficient peaks at the design wavelength of CGH which is 700 nm as expected.

\begin{figure*}[!htb]
  \centering
    \includegraphics[width=180 mm]{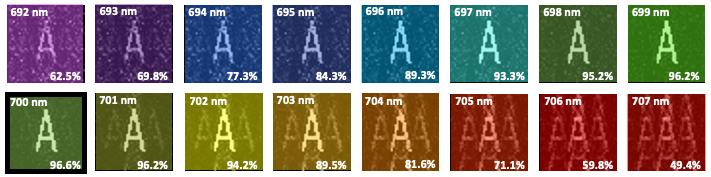}
    
  \caption{Variation of holographic images under different illumination wavelengths of light (692 nm - 707 nm) obtained with the hologram in Fig. \ref{wavelengthselection}b. The design wavelength of the holographic image A is 700 nm.}
  \label{Subimages}
\end{figure*}

In Fig. \ref{Subimages} we present holographic images of letter A under different illumination wavelengths of light between 692 nm and 707 nm. As seen here beyond and before the design wavelength of the CGH (700 nm), holographic images lose clear shape of letter A, and diffraction orders appear in the images. Especially, beyond the design wavelength of the CGH between 701 nm and 707 nm, diffraction orders in the holographic images are dominant. The diffraction orders can be eliminated by selection of longer distance between the hologram plane and the image plane. In that case correlation coefficient at the illumination wavelength will change, and the same hologram may not form the image of letter A. These holographic images are acquired for the selected design parameters as expected, and by considering updated design parameters the hologram can be re-designed.

In Fig. \ref{wavelengthselection}c we see that correlation value is less than 10\% when the illumination wavelength shifts 30 nm from the design wavelength of the CGH. With this figure, we understand that the intensity image of letter A is drastically distorted when the wavelength of light shows a 30 nm difference from the design wavelength of the optimized CGH. This wavelength shift is strongly affected by the distance between the CGH plane and the image plane, size of the CGH plane and the image plane, size of each pixel in the CGH plane and the image plane, and wavelength of light source. As a result, we conclude that when the design wavelengths of a color holographic image are 30 nm apart there is no overlap between the images. Therefore, a true multi-color hologram should work beyond 30 nm bandwidth.

\begin{figure*}[!htb]
  
    
    \includegraphics[width=180mm]{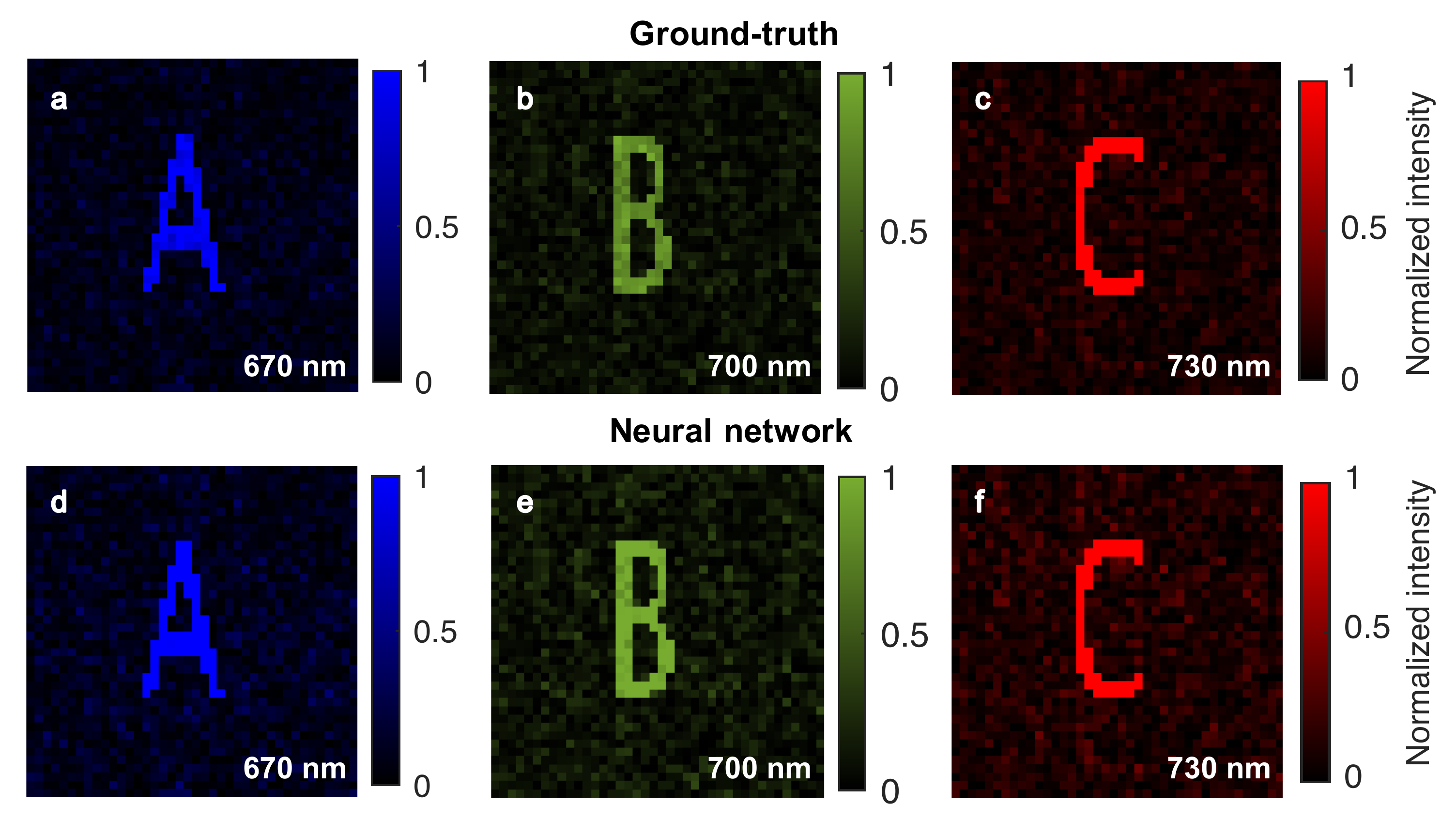}

  \caption{Generation of a holographic color image at a single observation plane. Ground-truth images of (a) letter A, (b) letter B, and (c) letter C. The DOENet based holographic images of (d) letter A, (e) letter B, and (f) letter C.}
  \label{ColorSingleObservationPlane}
\end{figure*}

\begin{figure}[!htb]
  
    
    \includegraphics[width=90 mm]{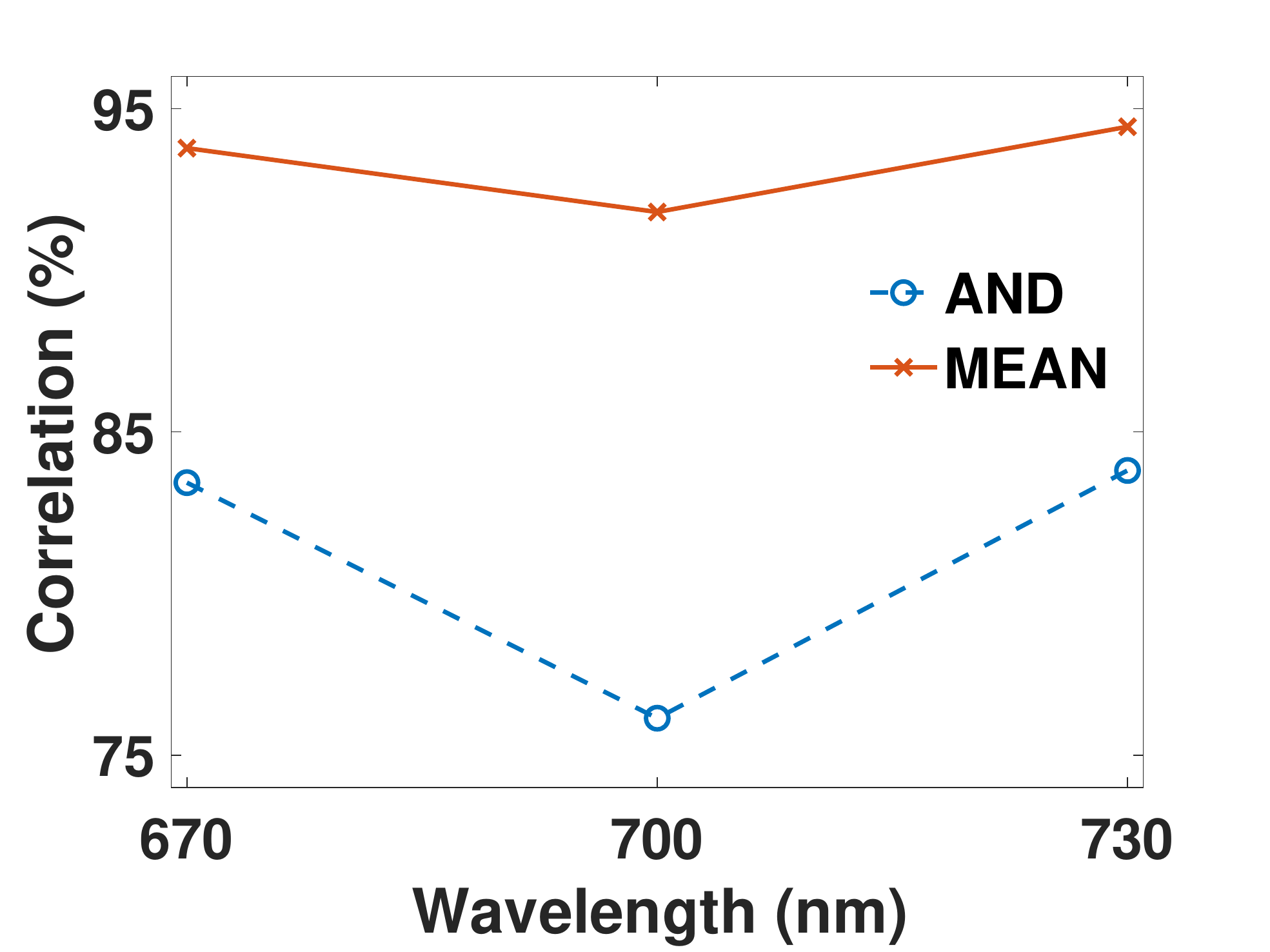}

  \caption{Change of correlation coefficients at design wavelengths of the holographic color image with two logic operations: AND and MEAN.}
  \label{logicoperations}
\end{figure}

\begin{figure}[!htb]
    
    \includegraphics[width=90 mm]{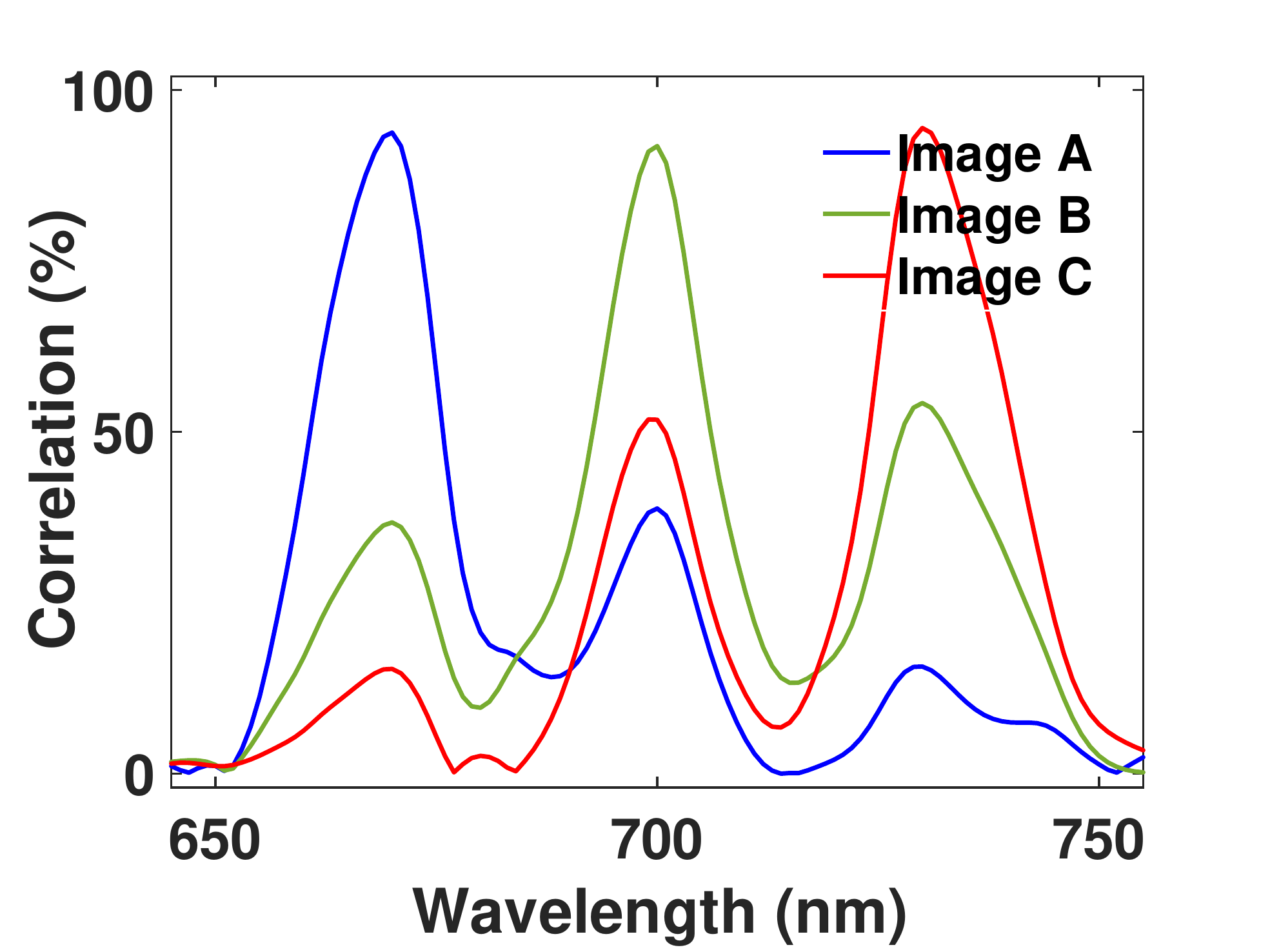}

  \caption{Correlation spectra of letter images A, B, and C when illumination wavelength spans between 650 nm and 750 nm.}
  \label{crosstalk}
\end{figure}

At first, we tuned thickness distribution of a CGH for generation of a holographic color image at a single observation plane. The wavelengths of the color image are 670 nm for alphabet image of A, 700 nm for alphabet image of B, and 730 nm for alphabet image of C. The transmitted images from the hologram plane are projected onto a white screen 350 $\mu$m away from surface of the CGH. The thickness optimization of the hologram was performed for four fully scannings of all the hologram pixels which last 3.6 hours with the local search optimization algorithm. The holographic images of letters are presented in Fig. \ref{ColorSingleObservationPlane}a-c. The intensity values on pixels of the letter figures are higher than background, and we observe formation of three holographic images with high contrast. For generation of these images, we use the equation of correlation coefficient in Eq. 4 as a cost function with MEAN logic operation. When AND logic operation is employed, the holographic images present a mean correlation of 81.1\% (Fig. \ref{logicoperations}). With MEAN logic operation we received a higher mean correlation value of 93.3\% due to increasing mean of correlation coefficient calculated with correlation coefficients at all design wavelengths. The images of letters A, B, and C show correlation coefficients of 93.8\%, 91.8\%, and 94.4\% in Fig. \ref{logicoperations}, respectively.

When the designed hologram is illuminated by wavelength of light at 670 nm, we see low correlation coefficients of 36.8\% and 15.3\% between letters A and B and letters A and C on the image plane of letter A, respectively (Fig. \ref{crosstalk}). However, cross-talk in Fig. \ref{ColorSingleObservationPlane}a is not visible even though there is a correlation of 36.8\% between letters A and B, and we observe only formation of letter A without formation of letters B and C. The ideal, uniform, and binary holographic image of letter A shows correlation coefficients of 34.8\% and 12.0\% with letters B and C, respectively. These high values are due to the fact that a great portion of the intensity image pixels is formed with zero intensity values. Moreover, the spatial similarity of the letters leads to a correlation which is expected. Therefore, correlation coefficients of 36.8\% and 15.3\% between letters A and B and letters A and C are not surprising, respectively. A similar situation occurs for letters A and C when illumination wavelength of light is 700 nm. There is no sign for formation of letters A and C in Fig. \ref{ColorSingleObservationPlane}b, but there are 51.8\% and 38.8\% correlation coefficients between letters B and A and letters B and C in Fig. \ref{crosstalk}, respectively. The ideal, uniform, and binary holographic image of letter B shows a correlation coefficient of 44.7\% with letter C. Correlation coefficients between letters C and A and C and B are 54.2\% and 15.7\% in the same figure when the same hologram is illuminated with light at a wavelength of 730 nm, respectively. However, there is no sign for formation of letters A and B on the image plane of letter C in Fig. \ref{ColorSingleObservationPlane}c.

%We obtained a correlation difference of 9.5\% between the ideal and the generated holographic images for letters A and B and a correlation difference of 3.3\% the ideal and the generated holographic images for letters A and C, a single hologram performs superiors to form three holographic images simultaneously.

While tuning thickness profile of the hologram for holographic images of letters A, B, and C, we stored intensity images and holograms at each optimization attempt. With the collected data set we tuned weights of the DOENet to reconstruct object information from an intensity color holographic image. The DOENet yields 99.7\% accuracies with the training and the validation data sets. We reconstruct a hologram with the DOENet which shows a correlation coefficient of 99.9\% with the ground-truth hologram. The reconstructed hologram by the DOENet is high-fidelity object information and obtained within two seconds. When the reconstructed hologram is illuminated with a light source at 670 nm, 700 nm, and 730 nm we see holographic three images as presented in Fig. \ref{ColorSingleObservationPlane}d-f. These three holographic images of letters are obtained with the reconstructed hologram by using ASM and Eqs. (1-3). These holographic images present correlation coefficients of 99.9\% with the ground-truth images. By using the DOENet, we reconstruct a hologram structure that provides a color image at a single observation plane within two seconds. This holographic structure behaves color filter to simultaneously achieve a full-color holographic image with no cross-talk. Moreover, we see twin image-free holographic intensity images due to elimination of recovery of phase and amplitude information from the holographic intensity images.

\begin{figure*}[!htb]
  
    
    \includegraphics[width=180mm]{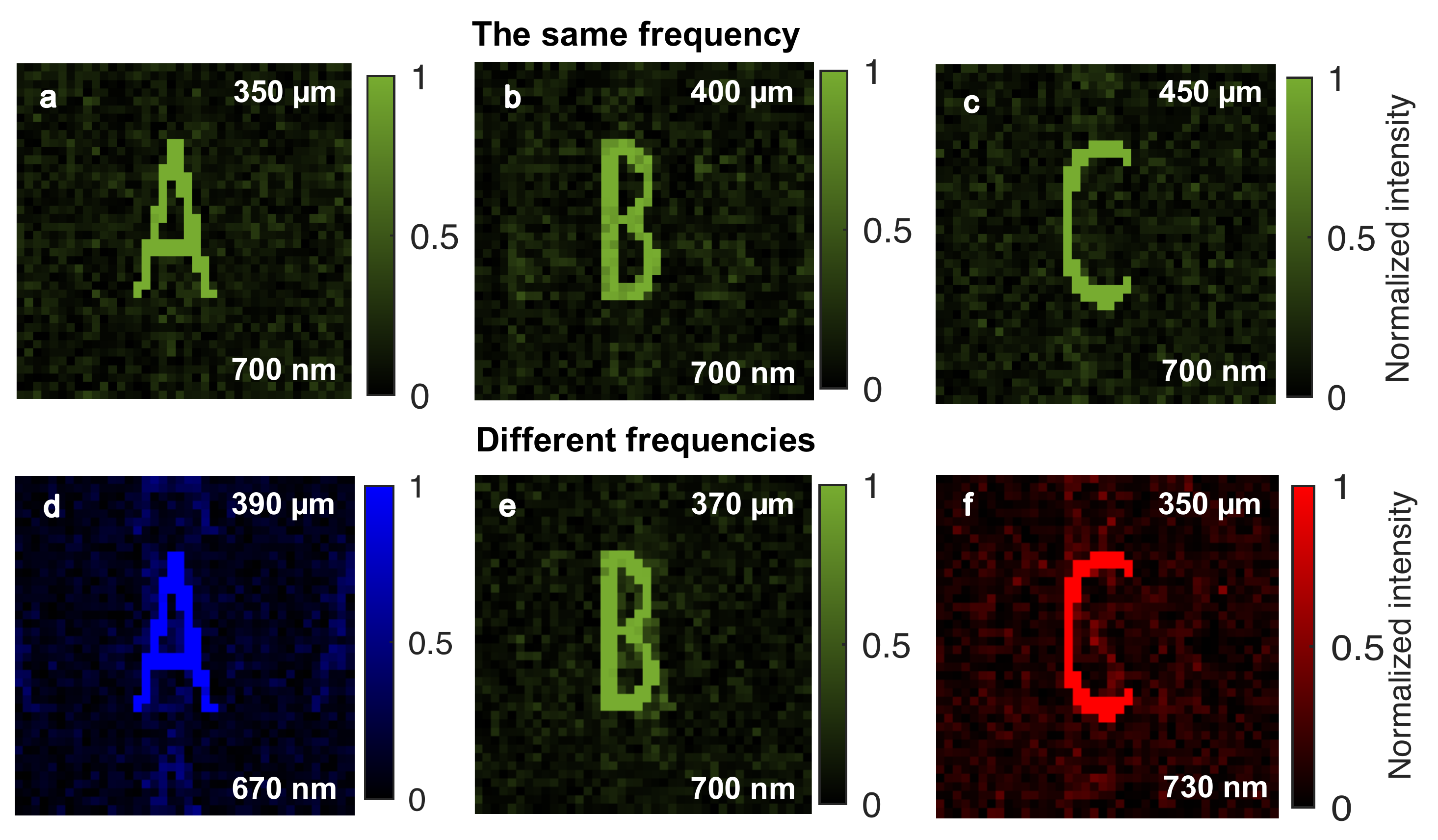}

  \caption{The DOENet based holographic image reconstruction. Monochrome holographic images at a wavelength of 700 nm and different observation planes: (a) Letter A at observation plane of 350 $\mu$m, (b) Letter B at observation plane of 400 $\mu$m, (c) Letter C at observation plane of 450 $\mu$m. Color holographic image at different wavelengths of light and observation planes: (d) Letter A at wavelength of 670 nm and observation plane of 390 $\mu$m, (e) Letter B at wavelength of 700 nm and observation plane of 370 $\mu$m, (f) Letter C at wavelength of 730 nm and observation plane of 350 $\mu$m.}
  \label{OnlyDOENet}
\end{figure*}

Next, we reconstructed a hologram structure that images three different figures forming at different observation planes (Fig. \ref{OnlyDOENet}a-c). The holographic images are arranged in order from letter A (350 $\mu$m), letter B (400 $\mu$m) to letter C (450 $\mu$m) where letter A is the nearest to the hologram plane and letter C is the farthest to the hologram plane. The DOENet shows accuracies of 99.7\% with the training and the validation data sets, and the reconstructed hologram with the DOENet has a correlation coefficient of 98.5\% with the ground-truth hologram. Later, the reconstructed hologram is illuminated with a light source emitting at a single wavelength of 700 nm. With ASM, we perform a numerical calculation for holographic intensity images seen at 700 nm and three observation planes. As seen in Fig. \ref{OnlyDOENet}a-c, we obtained clear and high contrast holographic images of letters formed at the same observation planes. These images are almost the same as the ground-truth images and present correlation coefficients of 99.9\% with the ground-truth images.

Lastly, the DOENet provides us reconstruction of a hologram that generates a color holographic image at three observation planes (Fig. \ref{OnlyDOENet}d-f). The letters A, B, and C are seen 390 $\mu$m, 370 $\mu$m, and 350 $\mu$m away from the hologram plane, respectively. Using the training data set of 51200 color images and holograms the DOENet yields accuracies of 99.4\% with the training and the validation data sets. The hologram designed with the DOENet has a correlation coefficient of 98.9\% with the ground-truth hologram. When the reconstructed hologram is illuminated with a light source emits at three wavelengths: 670 nm, 700 nm, and 730 nm, a color holographic image is seen at the aforementioned observation planes. The letter A is formed at 390 $\mu$m away from the hologram plane at a wavelength of 670 nm due to encoded spatial and spectral information into the reconstructed hologram (Fig. \ref{OnlyDOENet}d). In the meantime, we see formation of letter B at a wavelength of 700 nm and an observation plane of 370 $\mu$m (see Fig. \ref{OnlyDOENet}e). The same hologram images letter C at a wavelength of 730 nm and an observation plane of 350 $\mu$m as seen in Fig. \ref{OnlyDOENet}f. These holographic images present high correlation coefficients of 99.9\% with the ground-truth images.

The deep learning model used in this study decodes three-dimensional information of an object from an intensity-only recording. Our results qualitatively and quantitatively with correlation coefficient demonstrate effectiveness of the DOENet framework for formation of holographic images having different properties as wavelength, figure, and display plane. The DOENet does not expand a possible solution set that is constrained by physics. Instead, the DOENet finds the best solution within the data set considering desired intensity distribution. Compared to exhaustive search algorithms we create holograms by using the DOENet. The developed neural network architecture may boost design duration of a holographic image obtained by a hologram displayed on a spatial light modulator or a digital micro-mirror device. This work can be further improved to observe holographic images for multiple observers at oblique viewing circumference. Also if phase modulation capability of a hologram is extended with increasing number of thickness values in a hologram, clarity of holographic images improves. We believe the DOENet can be used to retrieve imaginary (phase) and real (intensity) parts of light from an intensity-only holographic image.

\section{Conclusion}

In this work using a single deep learning architecture the DOENet we demonstrate holographic image formation not only at different frequencies but also different observation planes. The DOENet enables us to obtain phase-only holographic structures to perform generation of holographic intensity images in two seconds. Moreover, thanks to data-driven statistical retrieval behavior of deep learning, we reconstruct hologram/object information from holographic images without requiring phase and amplitude information of recorded holographic intensity images. Compared to the iterative optimization methodologies our neural network produces holograms with several orders of magnitude faster and up to an accuracy of 99.9\%. The reconstructed holograms give superior quality reproduced intensity patterns. We believe our work inspires a variety of fields as biomedical sensing, information encryption, and atomic-level imaging that benefit from frequent accommodation of optical holographic images and object/hologram recovery.

\begin{acknowledgments}

This study is financially supported by The Scientific and Technological Research Council of Turkey (TUBITAK), grant no 118F075. PhD study of Alim Yolalmaz is supported by TUBITAK, with grant program of 2211-A. 
\end{acknowledgments}

\section*{Data Availability}
The data that support the findings of this study are available from the corresponding author upon reasonable request.

\bibliography{Library_Holography}% Produces the bibliography via BibTeX.

\end{document}